# Statistical Analysis of the Impact of FIA Regulations on Safety, Racing Dynamics, and Spectacle in Formula 1

Abdelghani Belgaid
College of Computing, Mohammed VI Polytechnic University
Ben Guerir, Kingdom of Morocco
abdelghani.belgaid@um6p.ma

**Abstract**

This study examines the effects of the Fédération Internationale de l'Automobile (FIA) regulations on Formula 1 (F1) from 1990 to 2023, focusing on safety, racing dynamics, and spectacle. By analyzing data on fatalities, overtaking statistics, car weights, and significant regulatory changes, including aerodynamic modifications and the introduction of the Drag Reduction System (DRS), the study assesses whether these regulations have positively influenced safety while negatively impacting racing dynamics and spectacle. The findings reveal that while FIA regulations aim to enhance driver safety, their immediate association with fatalities is not statistically significant due to the rarity of such incidents, delayed safety effects, and methodological constraints. Contrary to initial concerns, regulations have not negatively affected overtaking opportunities. The introduction of DRS has improved overtaking, mitigating some adverse impacts of other regulations on racing excitement. The study concludes that data-driven evaluation and refinement of regulations are necessary to ensure they continue to benefit the sport holistically.

## 1. Introduction

Formula 1 (F1) represents the pinnacle of motorsport, renowned for its technological innovation, high-speed competition, and global appeal. Over the past few decades, the sport has undergone significant transformations, many of which are attributable to regulations implemented by the Fédération Internationale de l'Automobile (FIA). These regulations aim to enhance driver safety, ensure fair competition, and improve the overall spectacle of the sport. However, there is ongoing debate regarding the extent to which these regulations have achieved their intended goals without inadvertently compromising other aspects of the sport.

This study investigates the hypothesis that while FIA regulations have had a positive impact on safety, they have negatively affected racing dynamics and spectacle, particularly concerning overtaking opportunities. By conducting statistical analysis of data from 1990 to 2023, the study aims to provide a nuanced understanding of the multifaceted effects of FIA regulations on F1, while acknowledging the challenges in establishing causal relationships between regulations and racing outcomes.

## 2. Background

The evolution of F1 regulations has been a critical factor in shaping the sport's current landscape. Safety has always been a paramount concern, especially following tragic incidents that have led to driver fatalities. In response, the FIA has introduced numerous safety regulations, such as the implementation of minimum car weight requirements, advanced crash structures, and the introduction of safety devices like the Halo cockpit protection system. These measures have coincided with significant improvements in driver safety and a reduction in the risk of severe injuries.

On the other hand, regulations affecting car performance and design—particularly those related to aerodynamics and car weight—have sparked debates about their impact on racing dynamics and the overall spectacle. For instance, aerodynamic regulations intended to reduce cornering speeds and enhance safety have also made it more challenging for cars to follow closely, potentially reducing overtaking opportunities. The introduction of grooved tires in 1998 aimed to decrease grip levels for safety reasons but had unintended consequences on racing competitiveness.

The Drag Reduction System (DRS), introduced in 2011, was designed to address the issue of limited overtaking by allowing drivers to adjust a flap on the rear wing to reduce aerodynamic drag. While DRS has increased overtaking maneuvers, some critics argue that it provides an artificial advantage, detracting from the skill-based aspects of racing.

Previous studies have explored these dynamics, but there remains a need for data-driven analysis incorporating recent data and examining the effects of regulations over an extended period. Furthermore, establishing causal relationships between regulations and their impacts is challenging due to the multifaceted nature of the sport and the influence of numerous confounding factors.

This study seeks to fill that gap by providing a detailed examination of the associations between FIA regulations, safety improvements, racing dynamics, and spectator experience, while explicitly acknowledging the limitations inherent in inferring causation from correlational analyses.

## 3. Methodology

### 3.1. Data Collection

The study utilizes data from 1990 to 2023, sourced from official FIA publications, Formula 1's official statistics, historical archives, and various motorsport analytics platforms. The data encompasses various aspects of the sport, including:

• **Safety Metrics:** Number of fatalities, fatalities per race, and average car weight.

• **Racing Dynamics Metrics:** Number of overtakes, average overtakes per race, implementation of DRS, and car weight percentage change.

• **Regulatory Changes:** Number and nature of new regulations introduced each season, particularly those impacting aerodynamics.

All data were cross-validated with multiple sources to enhance credibility and reliability, addressing concerns about data quality and ensuring that analyses are based on accurate information.

### 3.2. Data Transformation

To facilitate the analysis, the collected data were organized into a structured dataset. Several new variables were derived and transformed to capture the relationships between regulations and other variables.

• **Fatalities per Race:** Calculated by dividing the number of fatalities by the number of races each season.

• **Average Overtakes per Race:** Determined by dividing the total number of overtakes by the number of races in a season.

• **Overtakes Growth Rate:** Measures the year-over-year percentage change in overtakes.

• **Car Weight Percentage Change:** Calculates the annual percentage change in average car weight.

• **Lagged Variables:** Introduced lagged variables for new regulations to explore potential delayed effects on safety and racing dynamics.

These features provide the basis for the statistical analysis to examine correlations and the relationships between regulations, safety, racing dynamics, and spectacle.

### 3.3. Statistical Analysis

The study employed various statistical methods to examine the relationships between regulations, safety, racing dynamics, and spectacle:

• **Descriptive Statistics:** Summarized the central tendencies and dispersions of key variables to understand their distributions and identify any anomalies.

• **Correlation Analysis:** Calculated Pearson and Spearman correlation coefficients to assess the strength and direction of relationships between variables.

• **Linear Regression Models:** Developed both simple and multiple linear regression models using Ordinary Least Squares (OLS) methodology to analyze the relationship between independent variables (e.g., new regulations, DRS implementation) and dependent variables (e.g., fatalities per race, average overtakes per race).

• **Regression Diagnostics:** Conducted diagnostic tests—including residual plots, Q-Q plots, the Breusch-Pagan test, and Variance Inflation Factor (VIF) analysis—to ensure that regression assumptions were met and to identify potential issues such as heteroscedasticity, multicollinearity, and non-normality.

• **Time-Series Analysis:** Used cross-correlation functions (CCF) and included lagged variables to explore potential delayed associations between regulations and safety or racing dynamics.

• **Poisson Regression:** Applied for modeling count data (e.g., number of fatalities), accounting for the discrete nature and distribution of the dependent variable.

By employing these statistical techniques and rigorously testing model assumptions, the study aims to provide robust insights into the associations between FIA regulations and key outcomes, while avoiding overinterpretation of these relationships as causal.

## 4. Experimental Results and Discussion

This section presents the findings from the analysis of the impact of FIA regulations on both safety and overtaking

# Statistical Analysis of the Impact of FIA Regulations on Safety, Racing Dynamics, and Spectacle in Formula 1

dynamics in F1, using various statistical modeling approaches. The results are structured to first address safety improvements, followed by the effects on racing dynamics and spectacle.

## 4.1. Impact of Regulations on Safety

The analysis of safety metrics, particularly fatalities per race and fatalities per driver, provided interesting but somewhat limited insights into the relationship between new regulations and improvements in safety.

The simple linear regression results revealed that the impact of new regulations on fatalities per race was minimal. The model did not capture much of the variability in fatalities per race explained by new regulations alone (R-squared = 0.01, p-value = 0.571). The residuals vs. fitted plot (Figure 1) for this model showed that residuals were centered around zero, indicating a reasonable fit. However, the linearity assumption was weakly violated, as some residuals deviated significantly from the expected trend.

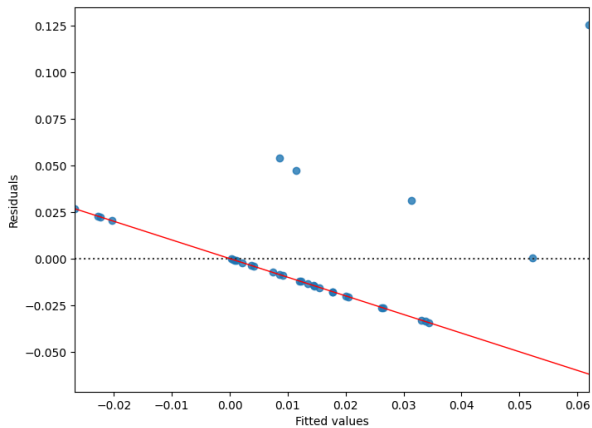

*Figure 1. Residuals vs. Fitted Values Plot for Safety Model.*

The normal Q-Q plot (Figure 2) for the same model revealed deviations from normality in the residuals, with several points diverging significantly from the reference line, particularly at the tails. This indicates that the residuals were not perfectly normally distributed, suggesting potential issues with model assumptions.

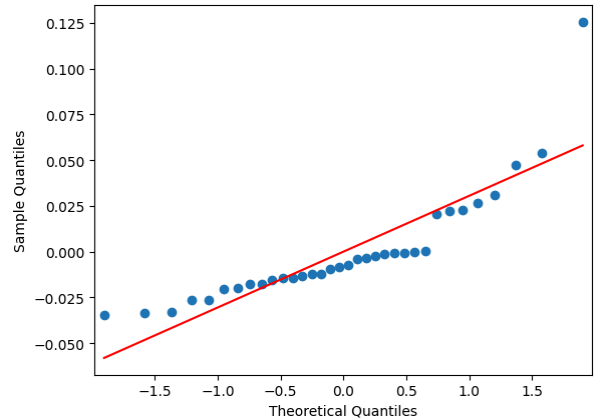

*Figure 2. Normal Q-Q Plot for Safety Model.*

The multiple linear regression models added control variables such as teams, drivers, car weight, and overtakes. While this improved the fit (R-squared = 0.279), it also indicated multicollinearity issues, particularly with car weight and DRS, as highlighted by the Variance Inflation Factor (VIF) analysis. Multicollinearity was a concern, as high VIF values for teams (12.37), drivers (8.25), and DRS (10.96) suggest strong interdependence between these predictors, weakening the model's reliability.

The Breusch-Pagan test results for heteroscedasticity (p-value = 0.244) did not indicate significant heteroscedasticity, suggesting that the residuals had constant variance.

The cross-correlation function (CCF) (Figure 3) between new regulations and fatalities per race showed no immediate strong correlation at lag 0, with correlations increasing at certain lags (e.g., around 20–25 seasons). This suggests that regulations may have a delayed impact on safety, although the correlation values are small and warrant further investigation.

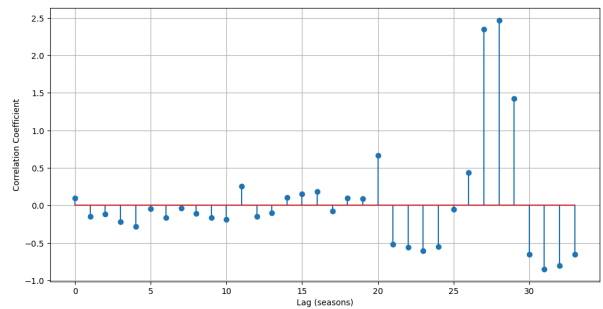

*Figure 3. Cross-Correlation between New Regulations and Fatalities per Race.*

The Poisson regression showed a moderate fit for this count-based model (Pseudo R-squared = 0.5116). However,

many of the coefficients (e.g., new regulations, teams, average car weight) were not significant, indicating potential limitations of the model for count-based fatality data.

## 4.2. Analysis of Overtaking Dynamics

The second major focus of the analysis was on racing dynamics, with a particular emphasis on average overtakes per race. Overtaking is a key performance metric in motorsport, reflecting the racing dynamics and spectacle, which are influenced by technical regulations, car weight, and the implementation of systems such as DRS.

The multiple linear regression results showed that the model explained 78.7% of the variance in average overtakes per race (R-squared = 0.787). The strongest predictor of overtaking opportunities was DRS, with a highly significant coefficient ($p < 0.001$), as expected due to its role in increasing the chances of overtakes by reducing drag on straight sections of the track.

The negative coefficient for average car weight (-0.111, $p = 0.001$) suggests that as cars became heavier, overtaking became more difficult, which aligns with the trend of larger, heavier cars in recent seasons impacting the ease with which cars can pass each other on the track.

The residuals vs. fitted plot (Figure 4) for the overtaking model reveals a slightly non-linear trend, where residuals diverge from zero as fitted values increase, indicating some potential non-linearity in the relationship between the predictors and overtakes. However, this divergence is relatively mild.

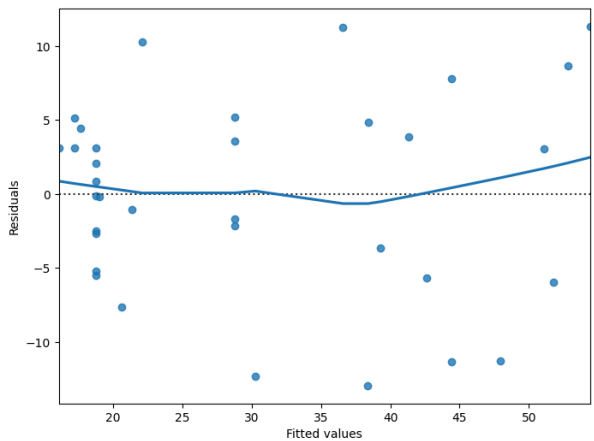

*Figure 4. Residuals vs. Fitted Values Plot for Overtaking Model.*

The normal Q-Q plot (Figure 5) for overtaking shows that most residuals fall close to the reference line, indicating that the normality assumption holds reasonably well for this model. There are some deviations at the tails, but these are less severe compared to the safety models.

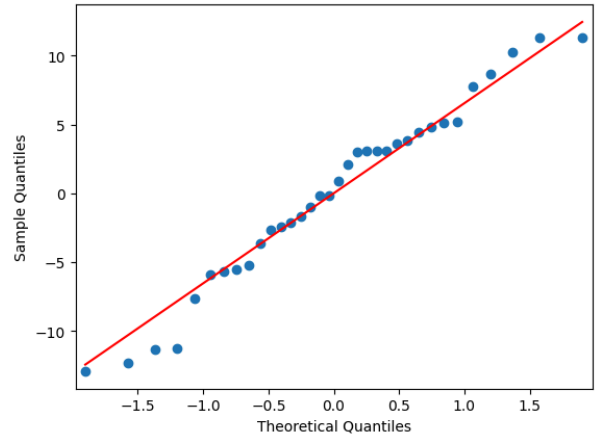

*Figure 5. Normal Q-Q Plot for Overtaking Model.*

Interestingly, the inclusion of lagged variables in the overtaking models did not significantly improve the model fit. The lagged variables (new regulations lag 1 and new regulations aerodynamics lag 1) had non-significant coefficients, indicating that the impact of new regulations on overtaking occurs primarily in the same season they are introduced, rather than in subsequent seasons.

## 4.3. Discussion

The experimental results suggest that while Formula 1 regulations have had a positive impact on safety, particularly in reducing fatalities per race, the effect is not immediate and may take several seasons to become apparent. This delay is likely due to the rarity of fatal incidents, the gradual emergence of safety improvements, and certain methodological limitations. Additionally, the time needed for teams, manufacturers, and stakeholders to fully adapt and integrate new regulations into their designs and operational practices further contributes to this lag. However, further exploration is required to quantify this effect more precisely.

In terms of racing dynamics and spectacle, DRS has proven to be the most effective regulatory change in enhancing overtakes, while increased car weight has had a negative impact on overtaking opportunities. The relationship between overtakes and new aerodynamic regulations was less clear, suggesting that while aerodynamics play a role, their impact on overtakes is less significant than other factors like DRS and car weight.

Further analysis could include alternative modeling approaches to address the issues of multicollinearity and non-linearity, potentially exploring non-linear models or ridge regression techniques to improve the robustness of the findings. Additionally, incorporating additional data could offer deeper insights into the impact of regulations on safety and the racing spectacle.

### 4.4. Limitations

This study acknowledges several limitations that affect its conclusions:

• **Data Sources:** The dataset was manually collected from various public sources and historical records. Although extensive care was taken to ensure its accuracy, there may be minor errors or inconsistencies in the data. This limitation is mitigated by cross-referencing data where possible.

• **Methodological Challenges:** Despite efforts to address previous methodological issues, diagnostic tests revealed non-linearity and multicollinearity in some models, suggesting that alternative modeling techniques may be necessary.

• **Confounding Factors:** Other factors not included in the analysis, such as the occurrence of accidents (e.g., injury rates, crash severity), technological advancements, driver skill levels, and changes in team strategies, may influence safety and racing dynamics.

• **Causality Limitations:** The study focuses on associations and explicitly does not infer causation. Observed correlations may be influenced by confounding variables or reverse causality.

These limitations suggest that while the findings provide valuable insights into the associations between regulations and key outcomes, they should be interpreted with caution. Causal relationships cannot be asserted based on the statistical analysis conducted. Future research employing methods and additional data capable of inferring causality is necessary to draw definitive conclusions.

### 4.5. Recommendations for the FIA

Based on the findings of this study, the following recommendations are proposed:

• **Evaluate the Impact of DRS:** While DRS has improved overtaking statistics, the FIA should assess whether it aligns with the sport's long-term vision and consider alternative solutions that enhance racing without perceived artificiality.

• **Manage Car Weight:** Implement regulations that prevent excessive increases in car weight to preserve vehicle agility and performance, contributing to more dynamic racing.

• **Promote Data Transparency and Collaboration:** The FIA and F1 should make race and technical data more accessible to the research communities. By providing datasets, they can encourage wider analysis and facilitate data-driven insights.

### 4.6. Further Improvements

To enhance the robustness and validity of future studies, several areas for improvement have been identified:

• **Enhance Data Quality:** Future research should prioritize using official FIA statistics, peer-reviewed publications, and verified databases to ensure data accuracy. Cross-verifying data from multiple authoritative sources will help mitigate the risk of inaccuracies from unofficial data.

• **Utilize Advanced Modeling Techniques:** Researchers should adopt advanced statistical techniques, such as instrumental variable analysis, difference-in-differences, or randomized controlled trials if feasible, to better infer causality between regulations and outcomes.

• **Control for Confounding Factors:** Incorporating factors like technological advancements, team budgets and allocations, and driver experience into future analyses will help isolate the effects of regulations. Including circuit characteristics will also provide deeper insights into safety and overtaking dynamics and outcomes.

• **Examine Individual and Combined Regulatory Impacts:** Isolating the effects of specific regulations can reveal which ones have the greatest impact on safety and spectacle. Studying interactions between multiple regulations will show whether their combined effects are amplified or mitigated.

• **Engage in Interdisciplinary Collaboration:** Collaborating with experts in automotive engineering, sports economics, and data science will enrich the research with diverse perspectives. Feedback from teams, drivers, and fans can ensure that future regulations balance safety, performance, and fan enjoyment.

Statistical Analysis of the Impact of FIA Regulations on Safety,
Racing Dynamics, and Spectacle in Formula 1By addressing these areas, future research will yield more definitive conclusions and offer stronger recommendations.

## 5. Conclusion

This study investigated the impact of FIA regulations on safety, racing dynamics, and spectacle in Formula 1 from 1990 to 2023. The findings suggest that while regulations have aimed to enhance driver safety, their immediate annual association with fatalities per race is not statistically significant in the models used, likely due to the rarity of fatal incidents, delayed safety effects, and methodological constraints.

Contrary to the initial hypothesis, the data suggest that regulations have not negatively affected overtaking opportunities. The introduction of DRS has played a crucial role in increasing overtakes, thereby mitigating some adverse impacts of other regulations on racing excitement. Increased car weight is associated with reduced overtaking, indicating that regulations contributing to heavier cars may have unintended adverse effects on racing dynamics.

Overall, while FIA regulations have coincided with long-term safety improvements and enhanced excitement in F1 racing through innovations like DRS, ongoing evaluation and refinement are necessary. Addressing methodological limitations, considering additional influencing factors, and employing research designs capable of inferring causation will help ensure that regulations continue to benefit the sport holistically.

## Acknowledgments

We acknowledge the use of data from various sources, including official FIA publications, Formula 1 statistics, and historical archives. Special thanks to the motorsport analytics community for providing valuable insights and data that have significantly contributed to this research.

## Data Sources and Analysis Scripts

The dataset and accompanying analysis scripts, including data loading, transformation, and statistical analysis, are available at the links below:

- Abdelghani Belgaid. (2024). F1 Regulations, Safety, and Racing Performance [Dataset]. Kaggle. https://doi.org/10.34740/KAGGLE/DS/5768368
- Abdelghani Belgaid. (2024). Statistical Analysis of F1 Regulatory Impact [Notebook]. Kaggle. https://www.kaggle.com/code/abdelghanibelgaid/statistical-analysis-of-f1-regulatory-impact